\documentstyle[emulateapj]{article}

\newcommand{\cha}{{\sl Chandra}}
\newcommand{\ein}{{\sl Einstein}}

\newcommand{\ros}{{\sl ROSAT}}
\newcommand{\etal}{et al.}
\newcommand{\logns}{{log$N$-log$S$}}

\slugcomment{Submitted to Astrophysical Journal Letters.}

\begin{document}

\title{
PROPERTIES OF THE \cha\ SOURCES IN M81}

\author{
Allyn F. Tennant\altaffilmark{1}, 
Kinwah Wu\altaffilmark{2},
Kajal K. Ghosh\altaffilmark{3},
Jeffery J. Kolodziejczak\altaffilmark{1}, and
Douglas A. Swartz\altaffilmark{4}
} 

\altaffiltext{1}{Space Science Department, NASA Marshall Space Flight Center, SD50, Huntsville, AL, USA} 
\altaffiltext{2}{School of Physics, University of Sydney, 2006, Australia; and MSSL, University College London, Holmbury St. Mary, Surrey, RH5 6NT, UK}
\altaffiltext{3}{NAS/NRC Senior Resident Research Associate, NASA Marshall Space Flight Center, SD50, Huntsville, AL, USA}
\altaffiltext{4}{Universities Space Research Association, NASA Marshall Space Flight Center, SD50, Huntsville, AL, USA}

\begin{abstract}
The Chandra X-ray Observatory obtained a 50-ks observation of the central region of M81 using the ACIS-S in imaging mode. The global properties of the 97 x-ray sources detected in the inner $8\arcmin.3 \times 8\arcmin.3$ field of M81 are examined.
Roughly half the sources are concentrated within the central bulge. The remainder are distributed throughout the disk with the brightest disk sources lying preferentially along spiral arms. 
The average hardness ratios of both bulge and disk sources are consistent with power law spectra of index $\Gamma$$\sim$1.6 indicative of a population of x-ray binaries. A group of much softer sources are also present.
The background source-subtracted \logns\ distribution of the disk follows a power law of index $\sim$-0.5 with no change in slope over three decades in flux. The \logns\ distribution of the bulge follows a similar shape but with a steeper slope above $\sim 4 \times 10^{37}$ ergs~s$^{-1}$.
There is unresolved x-ray flux from the bulge with a radial profile similar to that of the bulge sources. This unresolved flux is softer than the average of the bulge sources and extrapolating the bulge \logns\ distribution towards weaker sources can only account for 20\% of the unresolved flux. 
No strong time variability was observed for any source with the exception of one bright, soft source. 
\end{abstract}


\section{Introduction}

Among the best-studied galaxies beyond the Local Group is the nearby Sab spiral M81 (NGC 3031). 
With an inclination of 59$^{\circ}$, a dominant two-armed spiral pattern, and a well-defined central bulge, M81 is ideal for investigating formation and evolution of galaxies similar to our own.

Most x-ray studies of M81 have concentrated on the dominant source in the field, a low-luminosity AGN at its core. The only 
extensive examination of the entire M81 field at x-ray energies is that of Fabbiano (1988, hereafter F88) based on \ein\ observations. 
\ein\ detected 9 sources in M81 including the nucleus, all with x-ray luminosities exceeding $\sim 2 \times 10^{38}$ ergs~s$^{-1}$, with 5 of these located along spiral arms. We have identified an additional 9 sources within a $\sim$6$\arcmin$ radius of the nucleus from archival \ros\ observations.

In this first examination of a \cha\ X-ray Observatory observation of M81, we analyze the
global properties of the 97 x-ray sources detected in the inner $8\arcmin.3 \times 8\arcmin.3 $ field to a limiting 0.2-8 keV luminosity of $L\sim 3 \times 10^{36}$ ergs-s$^{-1}$. 
%
Details of the observation and the analysis methods are given in \S~\ref{s:data}.
Section~\ref{s:spatial} examines the spatial distribution of sources and 
identifies those sources spatially coincident with previously classified objects.
The hardness ratio distribution is presented in \S~\ref{s:spectral}
and the \logns\ distribution in \S~\ref{s:photometry}.
Section~\ref{s:temporal} discusses the luminosity evolution of sources common to the \ein\ (c.1980), \ros\ ($\sim$1990) and present \cha\ observations.
The results are summarized in \S~\ref{s:summary}.
%
Detailed analysis of individual sources are deferred to a future paper. 

\section{Observations} \label{s:data}

A 49926 second observation of a portion of the galaxy M81 was obtained
with the \cha\ Advanced CCD Imaging Spectrometer (ACIS, G. P. Garmire \etal, in preparation) on 2000 May 7.
The primary target, SN 1993J, was located near the nominal aimpoint on the
back-illuminated CCD S3 of the spectroscopy array (operating in imaging mode).
The nucleus of M81 lies 2.$\arcmin$79 from SN 1993J towards the center of S3 in this observation.
The observation was taken in faint time exposure mode at 3.241 s-frame$^{-1}$.
Standard \cha\ X-ray Center processing has applied aspect corrections and compensated for spacecraft dither. We selected the standard grade set and events in pulse invariant (PI) channels corresponding to $\sim 0.2$ to $8.0$ keV for analysis.

The majority of sources identified are within the $8\arcmin.3 \times 8\arcmin.3 $ field of S3.
To avoid difficulties with the intrinsic differences between
front- and back-illuminated devices we restrict our analysis to CCD S3.
Due to poor cosmic ray rejection for the central two columns of S3, 
all events with CHIPX equal to 512 or 513 were excluded from analysis.
Since these columns are dithered across the sky, their removal results
in a slight loss of sensitivity (5\%) in the central $16\arcsec$.
Pixels with centers less than $1\arcsec.25$ from the readout trail of
the nucleus were flagged and excluded from analysis
as were pixels containing less than 50\% sky exposure near the edge of
the CCD. 

Sources were located using a minimum variance estimator that compares the
data to the telescope Point Spread Function (PSF).
This method is most suited to finding point sources that roughly match the telescope's PSF.
The PSF was approximated by a circular gaussian with
the same width as the measured PSF at the aimpoint and increasing quadratically
with distance from the aimpoint to approximately match the core and off-axis
broadening of the true PSF.
We define the signal-to-noise ratio (S/N) to be the estimated source counts
divided by the 1$\sigma$ uncertainity in this number.
To set a minimum threshold, we considered the number of sources found
per S/N interval down to a minimum S/N of 1.0.
This distribution shows a clear gaussian noise peak at low S/N that indicates a single false detection (on S3) is to be expected for a minimum S/N of 3.0 and less than 0.1 false detections at a threshold of 3.5. We include all sources with S/N$>$3.0 in our analysis with the exception of the spectral analysis (\S~\ref{s:spectral}) and \logns\ analysis (\S~\ref{s:photometry}) for which only S/N$>$3.5 sources are considered. 

A second pass, designed to mitigate source confusion, was made through the data. For this purpose,
all pixels within 2$\sigma$ of each source position were flagged
and the source-finding process repeated. In this way, the variance
in the local background due to nearby sources is greatly reduced and otherwise confused sources isolated. An additional 6 sources, 4 with S/N$>$3.5, were detected.

The final source list contained 81 sources with S/N$\ge$3.5 with an additional
16 with 3.0$\le$S/N$<$3.5. 
These source positions are identified in the ACIS data shown in figure~1a and
superimposed 
on the second generation digital sky survery (DSS) image in figure~1b.

Small spatial regions centered on each source were then selected for further analysis.
The size of these spatial regions also varies quadratically with off-axis position and encompasses roughly the 95\% encircled energy radii of the PSF at 1.5 keV. No correction for the low background, $\sim$0.04 cts-pixel$^{-1}$, is made.
For spectral analysis (\S~\ref{s:spectral}), the total
counts within each source region were binned into soft (0.2~--~2.0 keV) and hard (2.0~--~8.0 keV) bands. 
Higher resolution spectra, at 16 PI channel binning out to $\sim$10 keV, were examined for spurious effects such as anomolously high background. No anomolies were encountered. 

Finally, we distinguish between the `bulge' and `disk' defined here as those regions internal and external, respectively, to a $4.\arcmin70 \times 2.\arcmin35$ ellipse centered on the nucleus with major axis position angle 149$^{\circ}$ (cf. figure~1). This ellipse corresponds to a circle in the plane of the galaxy of radius $2.\arcmin35$ ($\sim$2.5 kpc). There is extended, unresolved x-ray flux in the bulge that is above background and is not due to the PSF wings of the nucleus nor to other point sources. This component will be referred to as `excess bulge emission.'

\section{Morphology} \label{s:spatial}

The bulge contains 41 sources at a surface density of 5.4 sources-arcmin$^{-2}$
and the disk contains 56 sources (on S3) at a density of $\sim$1.0 source-arcmin$^{-2}$. Among the latter, 21 are located within $\pm$400 pc of spiral arms (as defined by Matonick \& Fesen (1997) by tracing the H$\alpha$ emission peak, their figure 31b). The surface density of spiral arm sources in our sample is slightly less than, but statistically consistent with, the surface density of disk sources not associated with spiral arms.

The surface density distribution of bulge sources, inclination-corrected and centered on the nucleus, can be described by an exponential radial distribution,
$\sigma(r) \propto e^{-r/r_o}$, with $r_o \sim 0.92\pm0.26$ kpc. 
After subtracting the background component and the contribution from the wings of the PSF of the nucleus, the surface brightness profile of the excess bulge emission can also be described by an exponential with scale length
$r_o \sim 0.83\pm0.20$ kpc for $0.\arcmin8\le r\le5.\arcmin5$. The excess bulge emission is much steeper at smaller radii though a residual contribution from the wings of the nuclear PSF cannot be definitively excluded. 
Beyond $r \sim 1$ kpc, these profiles are roughly consistent with the
$V$ (Georgiev \& Getov 1991), $g$ (Frei \etal\ 1996), and $B$ and $I$ 
(Elmegreen \& Elmegreen 1984) profiles suggesting the bulge x-ray emission follows the distribution of starlight.

A search of the literature for known objects spatially coincident with \cha\ sources found no matches within the bulge, with the exception of the nucleus.
Among the disk sources, 5 are located within $3\arcsec$ of supernova remnants (SNR; Matonick \& Fesen 1997) including SN~1993J and the bright \ein\ source X-6 (F88). All 5 SNR candidates lie on spiral arms as expected for young, massive star, Type II supernova progenitors. Source X-6 is a point source located within a 90-pc-diameter SNR (Matonick \& Fesen 1997). The high luminosity and hard spectrum (\S~\ref{s:spectral}) of X-6 is uncharacteristic of SNRs.

Six x-ray sources, all located within spiral arms, are within $10\arcsec$ of  
known HII regions, half of which are giant radio HII regions (Kaufman \etal\ 1987). One of these x-ray sources, though weak, has the hardest spectrum of the entire source sample (\S~\ref{s:spectral}). 

No x-ray sources were found to coincide with known globular clusters, novae, or stars.
Roughly 10\% of globular clusters are expected to 
contain accreting neutron stars detectable above our threshold and galaxies like M81 typically contain of order 200 globular clusters (Perelmuter \& Racine 1995). However, there are only 25 confirmed globular clusters in M81 (Perelmuter, Brodie \& Huchra, 1995) and only 4 within the S3 field. 

The nova rate in M81 is probably of order 20 year$^{-1}$ (A. W. Shafter, private communication). The 
x-ray signatures and light curves of novae are only sparsely documented (\"{O}gelman, Krautte \& Beuermann 1987)
making novae difficult to identify in our source sample. 

There are two sources, \ein\ source X-2 and a \ros-detected object, which are near stars or star clusters. X-2 is $1.\arcsec8$ distant from a star cluster with a spectrum consistent with G-type stars
(Sholukhova et al. 1998). Zickgraf \& Humphreys (1991) list the optical object as a probable foreground F-type dwarf. The \ros\ source is coincident with an 
undocumented $\sim$18-magnitude object visible in the Digital Sky Survey image.
Interestingly, both x-ray sources are transient and have been much brighter in the past (\S~\ref{s:temporal}). 

No known background or foreground objects were found coincident with the S3 \cha\ sources. Up to $\sim$25 background objects are expected in the S3 field above S/N~$=3.5$ based upon our analysis of the calibration field CRSS J0030.5+2618 (see also Brandt \etal\ 2000). This estimate ignores any obscurtion by M81.
Based on results of the \ein\  stellar survey (Topka \etal\ 1982), we estimate
roughly 0.4 F-~and G-stars could be detectable in our field  but these should all be extremely soft (Maggio \etal\ 1987, Hodgkin \& Pye 1994, Schmitt 1997). A soft spectrum was observed from 3 catalogued stars detected on CCD S2. 

\section{Hardness Ratio} \label{s:spectral}

The total source counts in the hard x-ray band (2.0~--~8.0 keV) is shown in figure~2 against the soft (0.2~--~2.0 keV) band  for the subset of sources with S/N$\ge$3.5 excluding the nucleus. There is no adjustment for pileup 
which tends to harden spectra of the brightest sources. 

Also shown are the hardness ratios of the sum of all the bulge sources, and of all the disk sources, the hardness ratio of the excess bulge emission (background and nuclear PSF subtracted), and of the typical background (scaled to a $2\arcmin \times 2\arcmin$ area). The nucleus and the bright soft source are omitted from the summed bulge point, and X-6 is omitted from the summed disk point as these sources are clearly spectroscopically atypical.

The line shown corresponds to an absorbed power law of spectral index $\Gamma = 1.6$, typical of x-ray binaries, with the Galactic absorbing column density of $N_H = 4 \times 10^{20}$ cm$^{-2}$ (Stark \etal\ 1992). Though indicative and useful to guide the eye, this canonical spectral shape is not a model fit result. Most sources fall near this curve. Notable exceptions are 
%
%
the soft sources at the lower right of figure~2. The softest bulge sources have no known associations but the 3 softest disk sources all lie on spiral arms. 
The softest source is also time-variable (\S~\ref{s:temporal}) and is the third brightest source in the entire sample. 
The hardness ratios for the summed bulge and summed disk spectra are also typical of x-ray binaries. The excess bulge emission is softer indicating some of the excess emission is from truly diffuse, hot, interstellar gas and not entirely from unresolved point sources.

\section{Log$N$-Log$S$} \label{s:photometry}

The number of sources above a limiting brightness, $N(>$$S)$, is shown against the 0.2~--~8.0 keV count rate, $S$, in figure~3. The 
nucleus is excluded as are sources with S/N$<$3.5.
Background sources were taken from the calibration field CRSS J0030.5+2618 (\S~\ref{s:spatial}), by applying our same analysis methods, and
have been subtracted after scaling to the fractional areas of the bulge and disk.

The curve shown in figure~3 represents the best power law fit to the disk distribution, $N =0.43S^{-0.50}$, over the entire range of $S$. Similarly, 
$N = 0.44 S^{-0.57}$ for the bulge source distribution for $S$$<$0.004 count-s$^{-1}$.
There is a break in the slope of the distribution of bulge sources above this point but no such break in the disk source distribution. Of the 10 disk sources with $S$$>$0.004 count-s$^{-1}$, 7 are coincident with spiral arms.

The luminosity of each source can be estimated by assuming the $\Gamma = 1.6$ power law spectral model of \S~\ref{s:spectral} 
and a distance of 3.6 Mpc to M81 (Freedman \etal\ 1994). 
For our 50-ks observation, therefore, $S=0.004$ count-s$^{-1}$ in the 0.2~--~8.0 keV band
corresponds to a luminosity $L_X = 3.7 \times 10^{37}$ ergs s$^{-1}$ and
the faintest source in the field corresponds to a luminosity of $\sim 3 \times 10^{36}$ ergs s$^{-1}$. 

Excluding the nucleus, the total
bulge luminosity is $L_X \sim 2.4 \times 10^{39}$ erg s$^{-1}$ 
of which 36\% is excess (unresolved) emission. The total luminosity of the disk sources is
$L_X \sim 3.9 \times 10^{39}$ erg s$^{-1}$. 
The luminosity of the nucleus is $L_X \sim 4 \times 10^{40}$ erg s$^{-1}$ based on spectral fits to the trailed image.
This is within the ASCA-observed range of luminosities (Ishisaki \etal\ 1996) and comparable to the BeppoSAX observed luminosity (Pellegrini \etal\ 2000).
The nucleus contributes approximately 86\% of the total luminosity in the $8\arcmin.3 \times 8\arcmin.3$ S3 field of view.

\section{Time Evolution} \label{s:temporal}

A simple test for time variability was made by binning the
light curves of all sources on 1000, 2000, and 4000 second intervals. Applying a $\chi^2$ test for the hypothesis of consistency with a mean value found only one clearly significant deviation on all three timescales. This source is also the softest source on S3 and the third brightest. It is present in at least 3 of 6 \ros\ HRI observations spanning 1993~-~1998 but is too close to the nucleus to be identified in any other previous x-ray observation.
 
M81 has been observed in x-rays at moderate spatial resolution by \ein\ in 1979 and by \ros\ over the period 1991--1998. There are 6 \ein\ sources (F88) within the S3 field of view and another 4 \ros-detected sources according to our analysis. 
\cha\ resolves six \ros\ source regions into two or more sources including \ein\ sources X-7 and X-10, and, of course, the nucleus. 

Long-term variability has been detected in
the \cha\ variable source, the nucleus, X-2 and the \ros-detected source coincident with an undocumented star-like object (\S~\ref{s:spatial}). Both of the latter sources are moderately weak in the present observations ($\sim$3 and $\sim 6 \times 10^{37}$ ergs-s$^{-1}$, respectively) but have been much brighter in the past.

The location of another \ein\ source, X-12, places it on the eastern edge of S3. There are no \cha\ sources within the portion of the 45$\arcsec$ error circle of this IPC source that falls on S3 and it is not detectable in the \ros\ observations. 

\section{Summary} \label{s:summary}

The global properties of the \cha\ sources identified in the central $8\arcmin.3 \times 8\arcmin.3$ region of M81 have been examined. There is a high density of sources located in the bulge of the galaxy. These sources, and the excess bulge x-ray emission, follow an exponential radial distribution about the nucleus with an $\sim$1 kpc scale length. This is similar to that of 
optical light and suggests 
the excess bulge emission is from unresolved point sources and that the point sources trace the old stellar bulge population. However, extrapolating the bulge \logns\ distribution towards weaker sources predicts $\sim$0.022 source counts-s$^{-1}$ from sources $<$0.001 counts-s$^{-1}$-source$^{-1}$. Thus, weak sources can only account for 20\% of the total excess bulge emission ($\sim$0.092 counts-s$^{-1}$). There remains a significant flux in the bulge unaccounted for by unresolved sources, particularly at lower energies, indicative of a truly diffuse, hot interstellar gas.
If all the excess bulge emission arises from a diffuse thermal bremsstrahlung at $kT \sim 0.2$ keV, uniformly distributed throughout the bulge, then there is 
$\sim 4 \times 10^6$ $M_{\odot}$ of hot gas.

The average hardness ratio of the \cha\ sources can be described by a power law of index $\Gamma=1.6$ with a Galactic absorbing column density of $N_H = 4 \times 10^{20}$ cm$^{-2}$. This is typical of x-ray binaries. Assuming thie canonical spectrum applies to all sources,
the background source subtracted \logns\ distribution of the disk sources follows an $N \propto S^{-0.50}$ profile extending to the most luminous sources at $L_X > 10^{39}$ ergs-s$^{-1}$. The distribution of bulge sources, however, shows a break at $L_X \sim 4 \times 10^{37}$ ergs-s$^{-1}$. Thus the bulge sources are predominantly low-mass x-ray binaries, consistent with an old bulge stellar population, while the disk has an additional population of much brighter sources.
Seven of the 10 brightest \cha\ disk sources lie within spiral arms, consistent with previous \ein\ observations (F88), but the entire sample of disk sources, down to the limiting \cha\ luminosity of $\sim 3 \times 10^{36}$ ergs-s$^{-1}$, are distributed throughout the disk without preference to spiral arms.
Some of these brighter sources may be black hole candidates with high-mass companions. 

\acknowledgements
We thank Martin Weisskopf for deriving the expression for the uncertainty in the source flux used in the source-finding algorithm.
The digital sky survey image used in Figure 1b is from 
the ``Palomar Observatory - Space Telescope Science Institute Digital Sky Survey'' of the northern sky, based on scans of the Second Palomar Sky Survey,
and was produced under NASA Contract NAS5-2555.

\newpage
\section*{Figure caption}

\figcaption{
{\em Left:}
\cha\ ACIS-S image of the inner region of M81 in J2000 coordinates. North is up and east is to the right.
Each sky pixel is $0.492\arcsec$. The aimpoint is near SN 1993J. The readout trail, caused by photons from the bright nucleus hitting the detector during readout, has been removed. \cha\ sources are marked as light blue (S/N$>$3.5) or magenta (3$<$S/N$<$3.5) circles. \ein\ sources (Fabbiano 1988) are marked by red circles encompassing the approximate \ein\ positional uncertainties. The $4\arcmin.7 \times 2\arcmin.35$ ellipse denotes the boundary between `bulge' and `disk' as used in the text.
{\em Right:} $B$-band digital sky survey image of M81 with the locations of sources identified as in left panel.
}
\figcaption{
Hardness ratio of the 80 sources with S/N$>$3.5, excluding the nucleus, with $\sqrt{N}$ errors. Also included are the sum of all bulge sources ($B$) excluding the bright soft source and the nucleus; the sum of all disk sources ($D$) excluding \ein\ source X-6; the excess (unresolved) bulge emission ($\diamond$); and the average instrument background ($\ast$) over a $2\arcmin \times 2\arcmin$ region. The line denotes the hardness ratio of a typical x-ray binary; a Galactically-absorbed power law of index $\Gamma = 1.6$. 
}
\figcaption{
Background-source-subtracted \logns\ for bulge and disk sources. The background \logns\ distribution was taken from \cha\ calibration observation of 
CRSS J0030.5+2618 (see also Brandt \etal\ 2000) using the same analysis procedures used in this study of M81. The line represents the power law fit to the disk distribution, $N =0.43S^{-0.50}$.
}

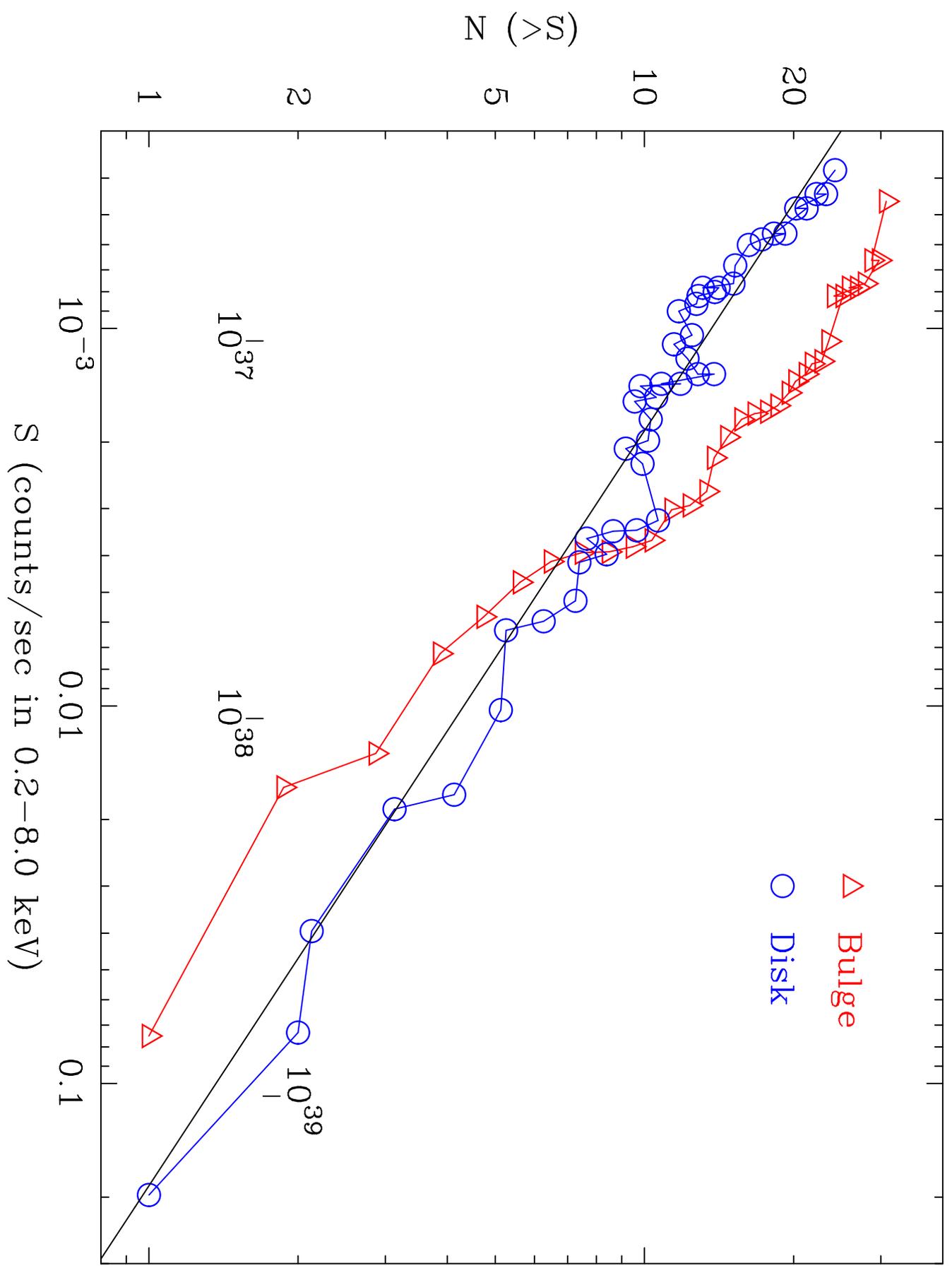
\end{document}